\documentclass[12pt]{article}

\usepackage{amsfonts,amsmath,amssymb,mathtools}
\usepackage{booktabs} % for tables
\usepackage{multirow} % for tables
\usepackage{graphicx}
\usepackage{xcolor} % for colored letters
\usepackage{hyperref} % for hyperlinks
\usepackage{lscape} % for landscape pages
\usepackage{natbib} % for references
\usepackage{enumerate}

\usepackage{bm}
\bmdefine{\btheta}{\theta}
\bmdefine{\brho}{\rho}
\bmdefine{\blambda}{\lambda}
\bmdefine{\bgamma}{\gamma}
\bmdefine{\bsigma}{\sigma}

\newcommand{\ihat}{\boldsymbol{\hat{\textbf{\i}}}} % i hat symbol
\newcommand{\jhat}{\boldsymbol{\hat{\textbf{\j}}}} % j hat symbol

\begin{document}

\def\spacingset#1{\renewcommand{\baselinestretch}%
{#1}\small\normalsize} \spacingset{1}

\thispagestyle{plain}
\begin{center}
\Large{\textbf{Bayesian Evidence Synthesis for Modeling SARS-CoV-2 Transmission}}
       
\vspace{0.4cm}
Anastasios Apsemidis$^{a*}$ and Nikolaos Demiris$^b$
\vspace{0.4cm}

\normalsize{$^{a*}$Department of Statistics, Athens University of Economics and Business, Athens, Greece, {\tt apsemidis@aueb.gr}} \\
\normalsize{$^b$Department of Statistics, Athens University of Economics and Business, Athens, Greece, {\tt nikos@aueb.gr}}
\end{center}

\bigskip
\begin{abstract}
The acute phase of the Covid-19 pandemic has made apparent the need for decision support based upon accurate epidemic modeling. This process is substantially hampered by under-reporting of cases and related data incompleteness issues. In this article we adopt the Bayesian paradigm and synthesize publicly available data via a discrete-time stochastic epidemic modeling framework. The models allow for estimating the total number of infections while accounting for the endemic phase of the pandemic. We assess the prediction of the infection rate utilizing mobility information, notably the principal components of the mobility data. We evaluate variational Bayes in this context and find that Hamiltonian Monte Carlo offers a robust inference alternative for such models. We elaborate upon vector analysis of the epidemic dynamics, thus enriching the traditional tools used for decision making. In particular, we show how certain 2-dimensional plots on the phase plane may yield intuitive information regarding the speed and the type of transmission dynamics. We investigate the potential of a two-stage analysis as a consequence of cutting feedback, for inference on certain functionals of the model parameters. Finally, we show that a point mass on critical parameters is overly restrictive and investigate informative priors as a suitable alternative.
\end{abstract}

\noindent%
\vfill

\newpage
\spacingset{1.9}
\section{Introduction}
\label{sec:intro}

The Covid-19 pandemic was initiated at Wuhan, China in late 2019 and is caused by the spread of the SARS-CoV-2 virus. The exact burden of the pandemic remains unknown and disease severity is highly variable with symptoms ranging from none or low fever to more serious such as chronic or death (see for instance \citealp{guan2020clinical}). The high transmissibility of the disease (see \citealp{flaxman2020estimating} and \citealp{tang2020estimation} for early estimates) led to preventive measures being adopted, initially in the form of non-pharmaceutical interventions (NPIs). The need for monitoring the epidemic led to the development of a vast range of models that could, in principle, assist the decision making process on the effect of NPIs against the virus. \cite{hellewell2020feasibility} assess the effect of contact-tracing and isolation, while \cite{kucharski2020early} discuss the effect of travel restrictions. \cite{eikenberry2020mask} investigate the use of face masks for protection. For a review over the Covid-19 modeling literature the reader can refer to \cite{cao2022covid}.

Traditionally, epidemic models are fitted on the recorded number of cases for inference or prediction but in the case of Covid-19 it has become apparent that only an unknown proportion of the total cases is observed. This phenomenon occurs due to the large number of asymptomatic cases as well as infected individuals not being tested for a number of reasons. Thus, statisticians should build upon these partially observed data to estimate quantities that, by definition, depend on the total number of cases. The current paper is based upon a suitably tailored stochastic discrete-time model presented in Section \ref{s:cov19mods}

The main contributions of this article are the following. A new discrete-time stochastic epidemic framework is presented and fitted to publicly available data. The unobserved number of infectious and susceptible individuals are estimated and independently validated against external data. The marginal likelihood is analysed in detail, while a distinct type of variable selection procedure is used for predicting the infection rate using mobility information when no dimension reduction takes place. Alternatively, principal components are determined by a criterion regarding the distortion of the original variables. The dynamics of the proposed system are examined in the phase-plane form and potentially viewed as a tool for monitoring the effectiveness of interventions via a simple visual inspection of the model assumptions.

The remainder of the article is organised as follows. In Section \ref{s:cov19mods} the models considered in the article are presented along with  methods for predicting the infection rate and the total cases. In Section \ref{s:results}, we present the results of training the models on data from the United Kingdom (UK), Greece, and the United States of America (USA), the model determination procedure and predictive performance. In Section \ref{s:phasean} we present the phase-plane analysis and its applications and we conclude with a discussion in Section \ref{s:discuss}.

\section{Modeling framework}
\label{s:cov19mods}

In this section we describe the proposed methodology for epidemic modeling at the country or state level, a suitably tailored stochastic SEIR in discrete time. The model comprises of four states, namely: Susceptible, Exposed, Infectious and Removed, abbreviated as $S$-state, $E$-state, $I$-state and $R$-state respectively. Note that while we focus upon the application to the Covid-19 data the model is applicable to other communicable diseases.

\subsection{Stochastic discrete-time transmission model}

Let $\mathbf{D}$ be a random vector of daily deaths and $d$ its recorded realization. We model the mean number of daily deaths using a Negative Binomial distribution so that the likelihood reads
\begin{equation} \label{eq:mod1eqn1}
P(D_t=d_t \,|\, \mathcal{F}_{t-1}) = NB(d_t; \theta_t, \psi)
\end{equation}
where $\mathbb{E}[D_t \,|\, \mathcal{F}_{t-1}]=\theta_t$ and $Var[D_t \,|\, \mathcal{F}_{t-1}]=\theta_t+\displaystyle\frac{\theta_t^2}{\psi}$ with $(\theta_t,\psi) \in \mathbb{R}\times\mathbb{R}^+$. The distributional form in \ref{eq:mod1eqn1} and other scale-mixtures of the Poisson distribution are explored in Section \ref{s:results}.

We wish to infer disease transmissibility based upon the total number of cases, $C$, as opposed to the observed ones. For a sample of $n$ days the model reads
\begin{align}
\theta_t &= p_t \cdot \sum_{k=1}^{t-1}\pi_{t-k} \cdot C_k \,, \quad \mathrm{for} \quad t=2,...,n \label{eq:mod1eqn2} \\  
C_t &= \lambda_{t-1-h}\frac{S_{t-1-h}I_{t-1-h}}{N} \,, \quad \mathrm{for} \quad t=\tau+h+1,...,n-1 \label{eq:mod1eqn3}
\end{align}
where $p_t$ is the Infection Fatality Ratio (IFR) at time $t$, i.e. the probability of death for a given case that occurred at time $s<t$, $C_t$ denotes the total cases at time $t$, $\lambda_t$ the infection rate and $N$ the population size. The length of the exposed period, whence an individual is infected but not infectious, is $h$ and $\tau$ is the infectious period with both assumed to be known from previous studies. We assume a piecewise constant IFR while the density of the time from infection to death is discretized using $\pi_s=\int\limits_{s-0.5}^{s+0.5}\pi(t)dt, s=2,...,n-1$ with $\pi_1=\int\limits_{0}^{1.5}\pi(t)dt$. The infection rate is piecewise constant in time with $J-1$ change-points at times $u_j$, i.e. $\lambda_t = \lambda_{(j+1)} \cdot I\big(t \in [u_j,u_{j+1}-1]\big)$, for $j=0,...,J-1$ with $u_0=1$ and $u_J=n-h-1$. The number of susceptibles $S_t$ and the active set of infectives $I_t$ are updated at each time point via
\begin{align}
S_t &= S_{t-1}-C_t \label{eq:mod1eqn4} \\
I_t &= \sum_{k=0}^{\tau-1} C_{t-k} \label{eq:mod1eqn5}
\end{align}
for $t=\tau,...,n-h-2$.
An important quantity in the epidemiology of infectious diseases is the basic reproduction number $R_0$, typically interpreted as the number of secondary infections generated by an infectious individual in a large susceptible population. The effective reproduction number at day $t$, $R_t$, is estimated using $R_t=\lambda_t \cdot \tau \cdot S_t/N$ and when $R_t>1$ ($<1$) the epidemic is (de-)escalating. The description of the model is completed by the $R$-state containing the number of individuals that recovered or died:
\begin{equation} \label{eq:mod1eqn6}
R_t^{(s)} = \sum_{i=1}^{t-\tau} C_i
\end{equation}
for $t=\tau,...,n-h-2$. 
Equations (\ref{eq:mod1eqn1}) -- (\ref{eq:mod1eqn6}) define the basic SEIR model which may be used for describing the acute phase of the pandemic and we now describe its extension which may be more appropriate for the endemic phase.

\subsubsection{Incorporating vaccination and demography}

A turning point of the pandemic was the introduction of the vaccine, which played a vital role in mitigating its influence. Vaccination is added in our model by assuming that after the first dose each individual remained susceptible for two weeks and then they became immune with probability $a_1$. After three more weeks, roughly corresponding to the time the second dose was given in many countries, we assume that the probability of immunity was raised to $a_1+a_2$. Thus, we incorporate the vaccine data by directly moving individuals to the $R$-state with a fixed probability. The updates of the susceptible equation (\ref{eq:mod1eqn4}) are now performed by
\begin{equation} \label{eq:mod2eqn4}
S_t = S_{t-1} - C_t - V_t
\end{equation}
where $V_t = a_1 \cdot \rho_{t-14} \cdot I\big(t \in [15,n-h-2]\big) + a_2 \cdot \rho_{t-35} \cdot I\big(t \in [36,n-h-2]\big)$ and $\rho_t$ is the number of vaccinations at time $t$. The removed individuals at time $t$ are now $R_t^{(s)} = \displaystyle\sum_{i=1}^{t-\tau} C_i + V_t$, while the infectious remain as in (\ref{eq:mod1eqn5}). We add further realism to the model by the inclusion of demography. This is not necessary when modeling acute outbreaks of short duration but over a period of three years one may wish to consider including births and deaths into the population dynamics. We assume that the number of births, $A$, is equal to the deaths (due to reasons other than Coronavirus) and update (\ref{eq:mod2eqn4}) as follows:
\begin{equation} \label{eq:mod3eqn4}
S_t = S_{t-1} - C_t - V_t + A \cdot (1-S_{t-1}/N)
\end{equation}
The $A$ term accounts for new births while the term $A\cdot S_{t-1}/N$ accounts for deaths. Furthermore, deaths from natural causes can occur inside the active set, so $I_t$ is updated by
\begin{equation} \label{eq:mod3eqn5}
I_t = \sum_{k=0}^{\tau-1} C_{t-k} - A\cdot I_{t-1}/N
\end{equation}
while the removed population is given by
\begin{equation} \label{eq:mod3eqn6}
R_t^{(s)} = \sum_{i=1}^{t-\tau} C_i + V_t - A\cdot R_{t-1}^{(s)}/N
\end{equation}
Note that newborn individuals are assumed to be susceptible and so we do not add births to $I_t$ or $R_t^{(s)}$. Thus, equations (\ref{eq:mod1eqn1}), (\ref{eq:mod1eqn2}), (\ref{eq:mod1eqn3}), (\ref{eq:mod3eqn4}), (\ref{eq:mod3eqn5}) and (\ref{eq:mod3eqn6}) define the SEIR model with vaccination and demography.

The $A$ term is informed by the data while $V_t$ is simply a transformation of data. The parameter $A$ is estimated using the population distribution across the published age groups as follows. Suppose that there exist $K$ age groups of length $g_i$ years, for $i=1,...,K$, and in each group belong $A_i$ people. Then, we assume that the newly born individuals are
\begin{equation} \label{eq:dembd}
A=\frac{A_1}{365 \cdot g_1}
\end{equation}

\subsection{Waning immunity and the SEIRS model}

The SEIR model presented thus far does not allow a transition from the $R$-state to the $S$-state. While a reasonable approximation at the early phase of the pandemic, re-infections cannot be ignored for longer time-horizons. Thus, we extend the SEIR model with vaccination and demography to the SEIRS setting where recovered individuals move to the susceptible state after $t^*$ days using $r_t=(1-p_t)\cdot\sum_{k=1}^{t-1}\pi_{t-k}^* \cdot C_k$, where $\pi_{t-k}^*$ is the discretized Gamma distribution of time from infection until recovery estimated in \cite{paul2021estimation}. Adding the lagged $r_t$ to $S_t$ in equation (\ref{eq:mod3eqn4}) via
\begin{equation} \label{eq:mod4eqn4}
S_t = S_{t-1} - C_t - V_t + A \cdot (1-S_{t-1}/N) + r_{t-t^*}
\end{equation}
allows recovered individuals to lose their immunity $t^*$ days after infection.

\section{Bayesian inference}

A key modelling decision relates to the selection of the death data as the basis for inference on the total cases and disease transmisibility. While this makes sense in terms of data quality, it implicitly precludes the use of case data for learning those parameters, effectively ``cutting feedback'' from the observed case data. This approach was introduced in \cite{spiegelhalter2007openbugs} and has been adopted in numerous studies in order to prevent data of low quality contaminating inference, see \cite{plummer2015cuts} for a review. We adopt this approach here, thus leading to a two-stage inference procedure, see for example Figure \ref{fig:cumpropgr} where the observed cases are used retrospectively for estimating quantities like the observed proportion.

\subsection{Computation and model determination}

The SEIR model described above is fitted under the Bayesian paradigm. The complexity of this non-linear hierarchical stochastic model implies that no analytical likelihood calculations exist. We resort to Hamiltonian Monte Carlo (HMC) sampling for posterior inference (see Supplementary Material D), notably the NUTS algorithm (see \citealp{hoffman2014no}), which is implemented in the \cite{rstan} probabilistic programming language. We have also extensively tested the Variational inference approximation used in Stan for comparison purposes and assessed its statistical and computational efficiency. In addition, we extensively tested the use of simulated annealing for inference. 

We test several structural assumptions of our modeling class using model selection procedures where appropriate. This was accomplished using information criteria but we also calculated the marginal likelihood (or model evidence) using Bridge sampling. Hence we also compared models via Bayes factors. The comparison of both the computational techniques and the distinct model structures is presented in Section \ref{s:results}.

\subsection{Prior specification and elicitation}

We assume independent priors on the $B$ infection fatality ratios $p_{(b)}$, the $J$ infection rates $\lambda_{(j)}$, the dispersion parameter $\psi$ and the initial value of total cases $C_1$ and fitted deaths $d_1$. The first $\tau+h$ values of $C_t$ are assumed equal and are given a $Gamma(2,0.0625)$ prior. The initial value of the susceptible population is then set to $S_1=N-C_1$. For each $\lambda$ between change-points $u_j$ and $u_{j+1}-1$ we assign a $LogNormal(0,1)$ prior. The dispersion parameter $\psi$ of the Negative Binomial is allocated a $Gamma(2, 0.125)$ prior. While these prior distributions are weakly informative, we also conducted sensitivity analyses to assess their effect as typically done in Bayesian robustness settings. We use a fixed $\tau=6$ infectious period (\citealp{cereda2020early}; \citealp{flaxman2020estimating}), and a fixed exposed period of $h=2$, since the mean incubation period is approximately 5 days (\citealp{lauer2020incubation}) and infectiousness starts approximately 2 days before the symptom onset (\citealp{he2020temporal}).

The prior on the IFR parameters $p_{(b)}$ requires particular caution since (i) it may not be informed by the outbreak data and (ii) it largely drives the scale of the estimated total cases and function(al)s thereof. For each interval of constant IFR, the prior was set to a strongly informative Gaussian distribution with standard deviation of $10^{-4}$ and mean computed as follows. First, we scaled the IFR published by the Centers for Disease Control and Prevention (CDC) for the 4 age groups 0-17, 18-39, 40-64 and 65+, $p^{(0)}_k$ ($k=1,...,4$), according to the age distribution of the recorded cases in the country under study. Then, we set the change-points by inspecting the observed country-specific IFR and use the mean IFR inside each time interval. Thus, if we denote by $c_{t,k}$ the cases at time $t$ for age group $k$ in a given country, the IFR is computed as $p_t^*=\sum_{k=1}^4 p^{(0)}_k \displaystyle\frac{c_{t,k}}{\sum_{i=1}^4 c_{t,i}}$ and we set the mean of the Gaussian prior to $\mathbb{E}[p_{(b)}]=\displaystyle\frac{1}{l_{b+1}-l_b}\sum_{t=l_b}^{l_{b+1}-1}p_t^*$. We also used this mean as a point mass prior for IFR but this approach was too rigid in some cases and we therefore opted for this strongly informative Gaussian prior.

\section{Results}
\label{s:results}

In this section, we analyse the pandemic data from the UK, Greece and the USA. Modeling the entire USA as a single entity may not be entirely appropriate due to the inherent heterogeneity stemming from both the large geographic area and population size so these results should be interpreted with caution. 

We present the results from the NUTS implementation of HMC. We also tried automatic differentiation variational inference (\citealp{kucukelbir2017automatic}) but, while this was faster in all case, it rarely gave reliable results in all but the simplest of models for short time horizons. In addition, we used Simulated Annealing (SA) in order to maximise the un-normalized posterior. Extensive searches by SA required less than 10 minutes when sampling took more than two days. However, the SA results were very sensitive to initial values while HMC appeared stable and robust. We therefore report HMC-based results due to statistical but not computational efficiency and retained HMC as our preferred algorithm for inference.

\subsection{Sources of evidence}

The publicly available data we leverage include the recorded cases and deaths, the number of vaccinated individuals, the age distribution of cases, the demographic births and deaths, the country population, the number of tests performed and quantities estimated at the beginning of the pandemic, namely the distribution of the infectious period, the exposed period, the serial interval and the time from infection until death. Below we outline the values we use for these quantities (also see Supplementary Material E).

For the UK, the cases and death data as well as the vaccination coverage and age distribution of cases are provided in the \citealt{phew} website. For the parameter $A$ we use \citealt{demuk} and calculate it using $g_1=5$ and $A_1=\displaystyle\frac{6.2 \cdot N_{UK}}{100}$. We use $N_{UK}=67886011$ as the 2020 \citealt{worluk}.

The Greek death data are obtained from the COVID-19 Data Repository of the Center for Systems Science and Engineering at Johns Hopkins University (\citealp{dong2020interactive}). The vaccine doses received are taken from Our World in Data (\citealp{mathieu2021global}), while the age distribution of cases and the daily tests performed were taken from \citealt{covgrgit}. The demographic births and deaths are obtained from the Hellenic Statistical Authority (\citealt{hsa}) which provides the population age groups and so we use equation (\ref{eq:dembd}) to calculate $A$ with $g_1=10$ and $A_1=1049839$. The population of Greece in 2020 is set to $N_{GR}=10816286$ following the HSA estimate. 

The USA death data are taken from the Johns Hopkins University while the cases and the vaccinations are obtained from Our World in Data. The population is set to be $N_{USA}=331230000$ and we calculate $A$ using $g_1=15$ and $A_1=\displaystyle\frac{18.51 \cdot N_{USA}}{100}$. For these we use the \texttt{statista.com} page, specifically the published \citealt{sta1us} and \citealt{sta2us} in 2020.

For the infection-to-death distribution we use the sum of the infection-to-onset of symptoms and onset-to-death distributions, which are Gamma with shapes 1.35 and 4.94 and rates 0.27 and 0.26 respectively, as in (\citealp{flaxman2020estimating}). The same source was used for the serial interval distribution as well. The $p_k^{(0)}$ IFR can be found in the \citealt{cdcifr} page. The probabilities of moving to the $R$-state are set to be $a_1=0.4$ and $a_2=0.1$.

\subsection{Epidemic parameters and functions thereof}

\begin{figure}
\begin{center}
\includegraphics[width=0.49\textwidth]{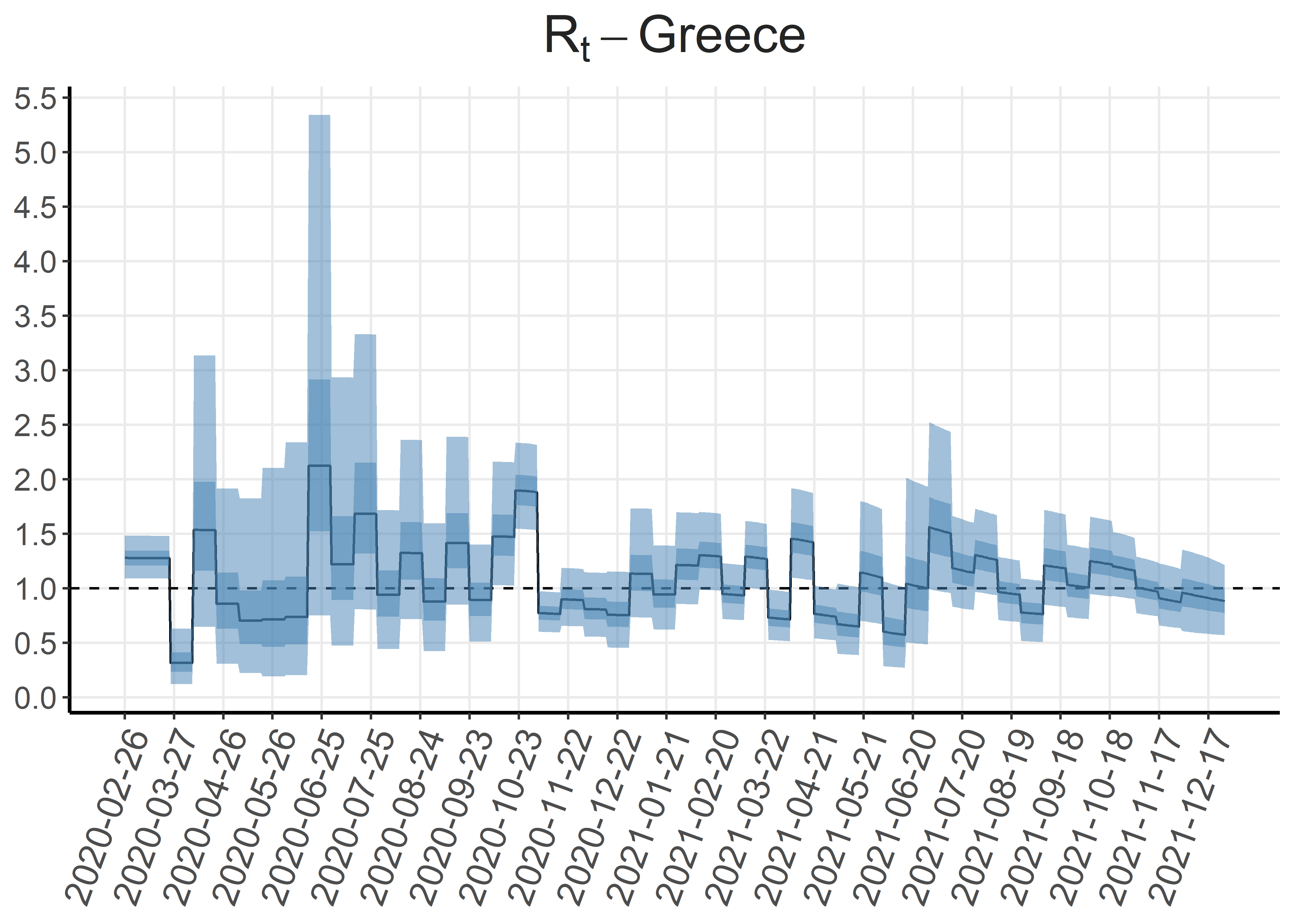}
\includegraphics[width=0.49\textwidth]{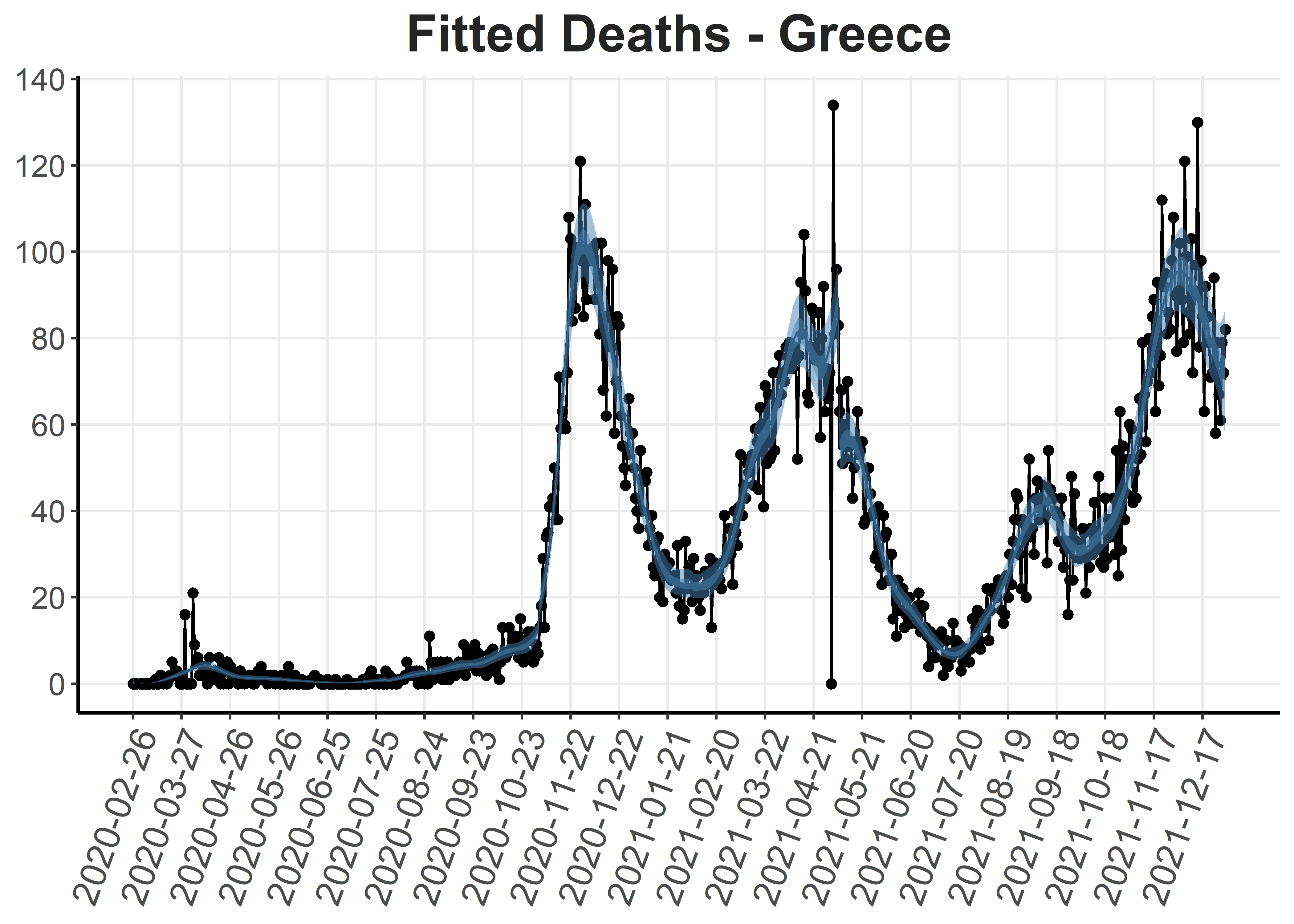}
\end{center}
\caption{Posterior estimates of $R_t$ (left) and $\theta_t$ (right) for Greece during the first two years of the epidemic. The median is depicted with a solid black line, along with 50\% and 95\% credible intervals.}
\label{fig:Rtthetadgr}
\end{figure}

We focus on the acute phase of the pandemic and fit the model to data from Greece covering the 26/2/2020 to 31/12/2021 period. We independently validate our estimates based on a large UK seroprevalence study.

\begin{figure}
\begin{center}
\includegraphics[width=0.49\textwidth]{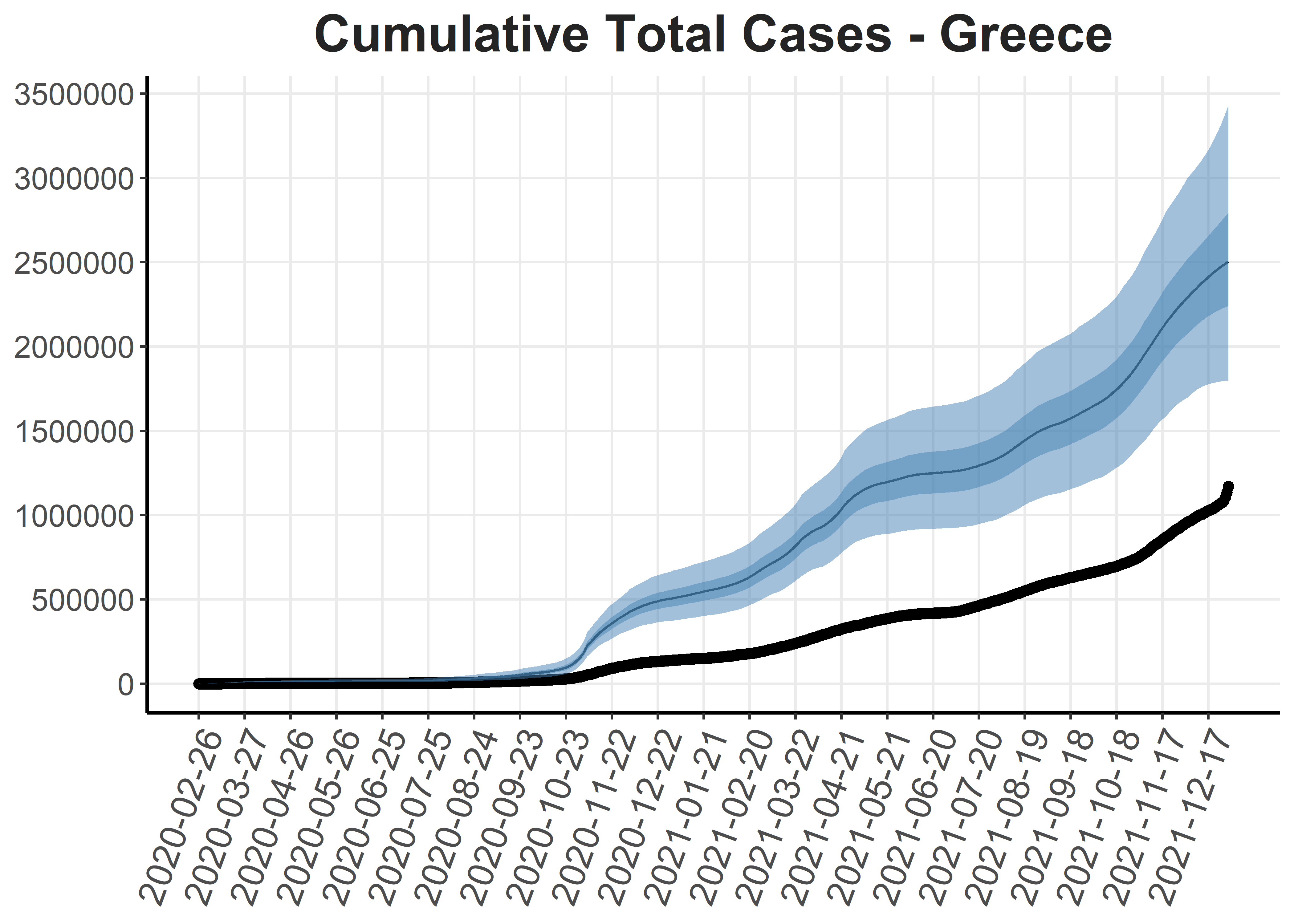}
\includegraphics[width=0.49\textwidth]{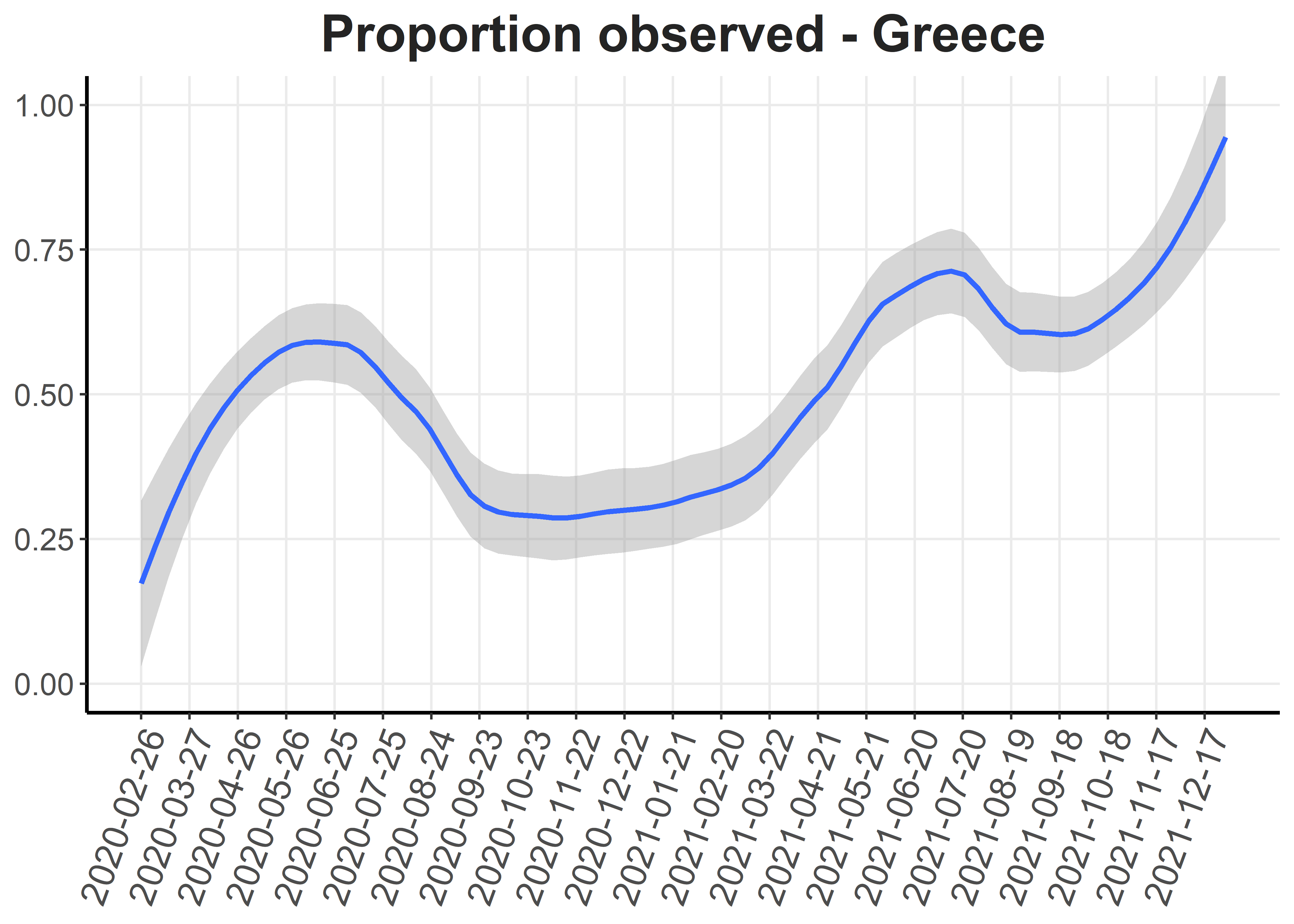}
\end{center}
\caption{Left: Cumulative total cases: posterior median (solid black line) with 50\% and 95\% credible intervals. Right: Proportion of observed cases smoothed by local regression.}
\label{fig:cumpropgr}
\end{figure}

In Figure \ref{fig:Rtthetadgr} the estimated reproduction number, $R_t$, and mean deaths $\theta_t$ are depicted. The fit to the death data is reassuring. The piecewise constant $R_t$ is scaled by the proportion of susceptible individuals as theory suggests. The total number of cases at day $t$, $C_t$, i.e. is the sum of the recorded and unrecorded ones are depicted in the left plot of Figure \ref{fig:cumpropgr} while the right panel contains the smoothed estimated proportion of observed cases, $c_t/C_t$. We estimate that the first million infections (10\% of the total population) was reached on April 2021 but observed eight months later on December 2021. The probability of recording a case initially was around 1/4 but then increased as tests and self-tests became widely available, closer to 3/4 of the total cases.

\begin{figure}
\begin{center}
\includegraphics[width=0.85\textwidth]{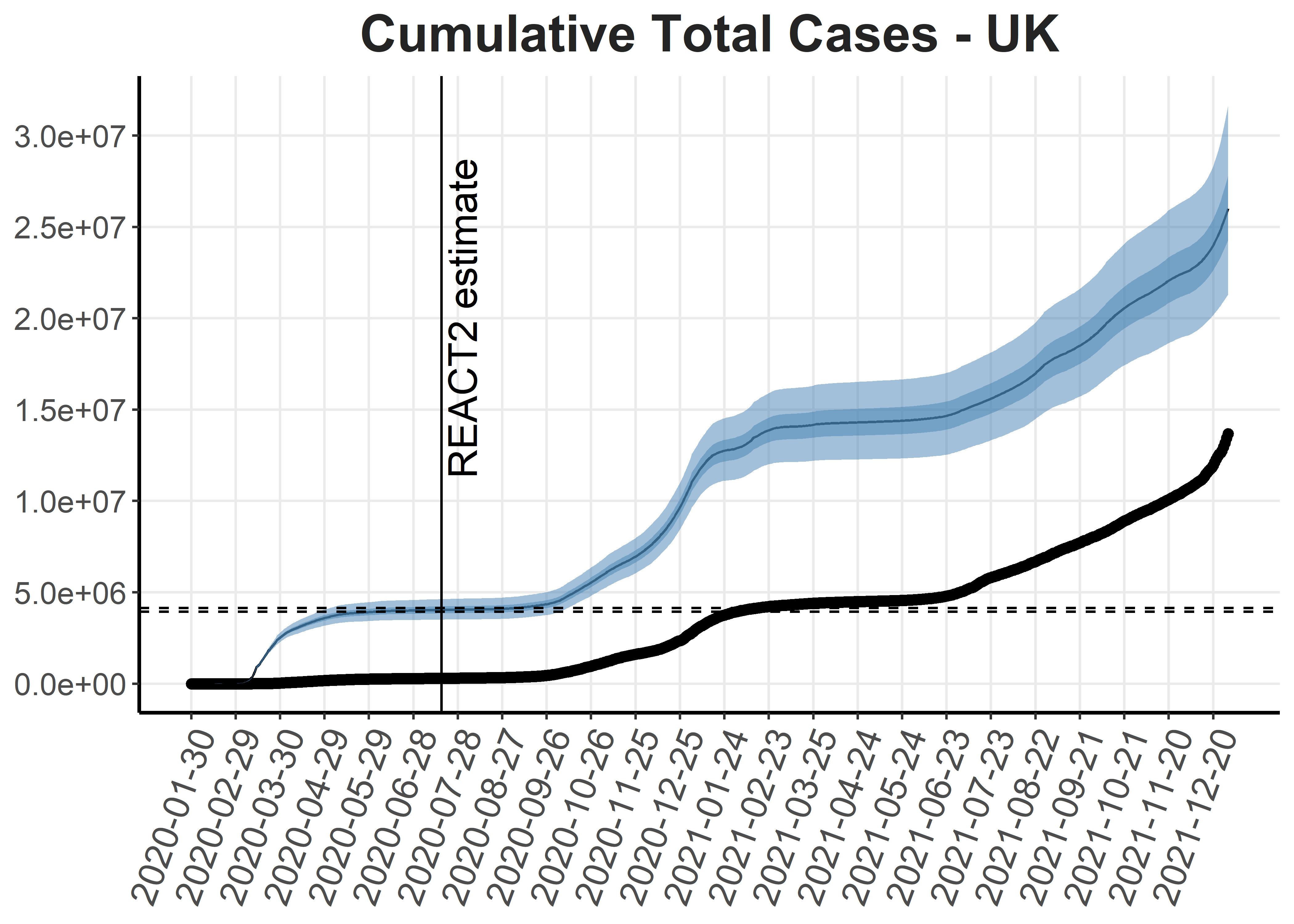}
\end{center}
\caption{Cumulative total cases for the UK. The time when the REACT-2 study estimated the total cases is noted as a vertical line, while the reported 95\% confidence interval is shown as horizontal dashed lines.}
\label{fig:ccasesreact}
\end{figure}

The external validity of our model was independently assessed by training the model on the UK death data and superimposing on Figure \ref{fig:ccasesreact} the estimated number of infections against the REACT study (\citet{ward2021sars}) which estimated an approximate 6\% prevalence on July 2020. It is apparent that our model estimate agrees well with this independent data source as desired. 

\subsection{Sensitivity Analysis and Model selection}

As part of the model comparison process we tested the initial values, the prior distributions, we added vaccination and demography to the model and tested seven scenarios where we removed the exposed period, vaccination or demography. In Table \ref{t:infcritmods1} we compare eight models for Greece: four SIR and four SEIR each with zero to two of the extensions ``vaccination'' and ``demography''. The comparison is made using the information criteria AIC, BIC, DIC, DIC where the effective number of parameters is set to half the variance of the deviance (referred to as DIC\textsubscript{2}) and WAIC. The time needed to fit each model is also displayed. In Supplementary Material F, we also compare the logarithm of the evidence $p(d_t)$ for each model using Bridge sampling and pairwise Bayes factors. Overall, the SIR with demography is selected by AIC and BIC, SEIR with vaccination or demography are preferred by DIC, the simple SIR is selected by DIC\textsubscript{2}, while the SEIR with vaccination and demography is preferred by WAIC. 

All the training times are similar and approximately equal to 2.5 days. The SEIR model with vaccination has the largest marginal likelihood, slightly higher than that of the SEIR with vaccination and demography while the interval estimates of the corresponding $\log p(d_t)$ show substantial overlap. Fitting the SEIRS model to the Greek data, allowing for re-infection 3 months after recovery ($3\cdot4\cdot7$ days) did not make any material difference to the SEIR version except for the last month, when the characteristics of the epidemic also change.

\begin{table}
\caption{Information criteria for the eight tested models and their training time measured in days for Greece. The results are rounded to 2 decimal digits.}
\label{t:infcritmods1}
\centering
\begin{tabular}{ c c c c c c c } 
\hline
& AIC & BIC & DIC & DIC\textsubscript{2} & WAIC & Time $(d)$ \\ \hline
SIR & 3989.98 & 4238.29 & 3969.52 & 3980.25 & 3980.82 & 2.67 \\
SIR.vacc & 3990.04 & 4238.35 & 3969.28 & 3984.23 & 3980.66 & 2.43 \\
SIR.dem & 3989.94 & 4238.25 & 3969.92 & 3986.48 & 3981.05 &  2.57 \\
SIR.vacc.dem & 3990.05 & 4238.36 & 3970.07 & 3985.86 & 3981.06 & 2.54 \\
SEIR & 3991.46 & 4239.76 & 3968.36 & 3984.83 & 3979.68 & 2.39 \\
SEIR.vacc & 3991.35 & 4239.67 & 3967.70 & 3982.07 & 3978.61 & 2.37 \\
SEIR.dem & 3991.35 & 4239.66 & 3967.70 & 3982.07 & 3978.61 & 2.50 \\
SEIR.vacc.dem & 3991.45 & 4239.76 & 3967.66 & 3983.79 & 3977.84 & 2.39 \\
\hline
\end{tabular}
\end{table}

The distributional form of the likelihood of the death data was tested as follows. Since the Negative Binomial is a scale mixture distribution of a Poisson with Gamma mixing rate, we also examined the standard Poisson distribution as well as scale mixtures with comparable Exponential and Log-Normal priors. Thus, this part of the model now reads $D_t \sim P(\theta_t)$ with $\theta_t \sim Exp(1/\mu_t)$ or $\theta_t \sim LN(m_t,s_t^2)$, where $m_t=\displaystyle\frac{\log(\mu_t^2)}{\displaystyle\sqrt(\mu_t^2+\sigma^2)}$ and $s_t^2=\log(1+\sigma^2/\mu_t^2)$. Then, $\mu_t=p_t\sum_{k=1}^{t-1}\pi_{t-k}C_k$ as we have with the Negative Binomial case. The Poisson-Exponential model (a special case of the Negative Binomial) fits the data well and reduces training time to a half. On the other hand, the Poisson-LogNormal struggled to converge and was unreliable.

\subsection{Transition to endemicity}

\begin{figure}
\begin{center}
\includegraphics[width=\textwidth]{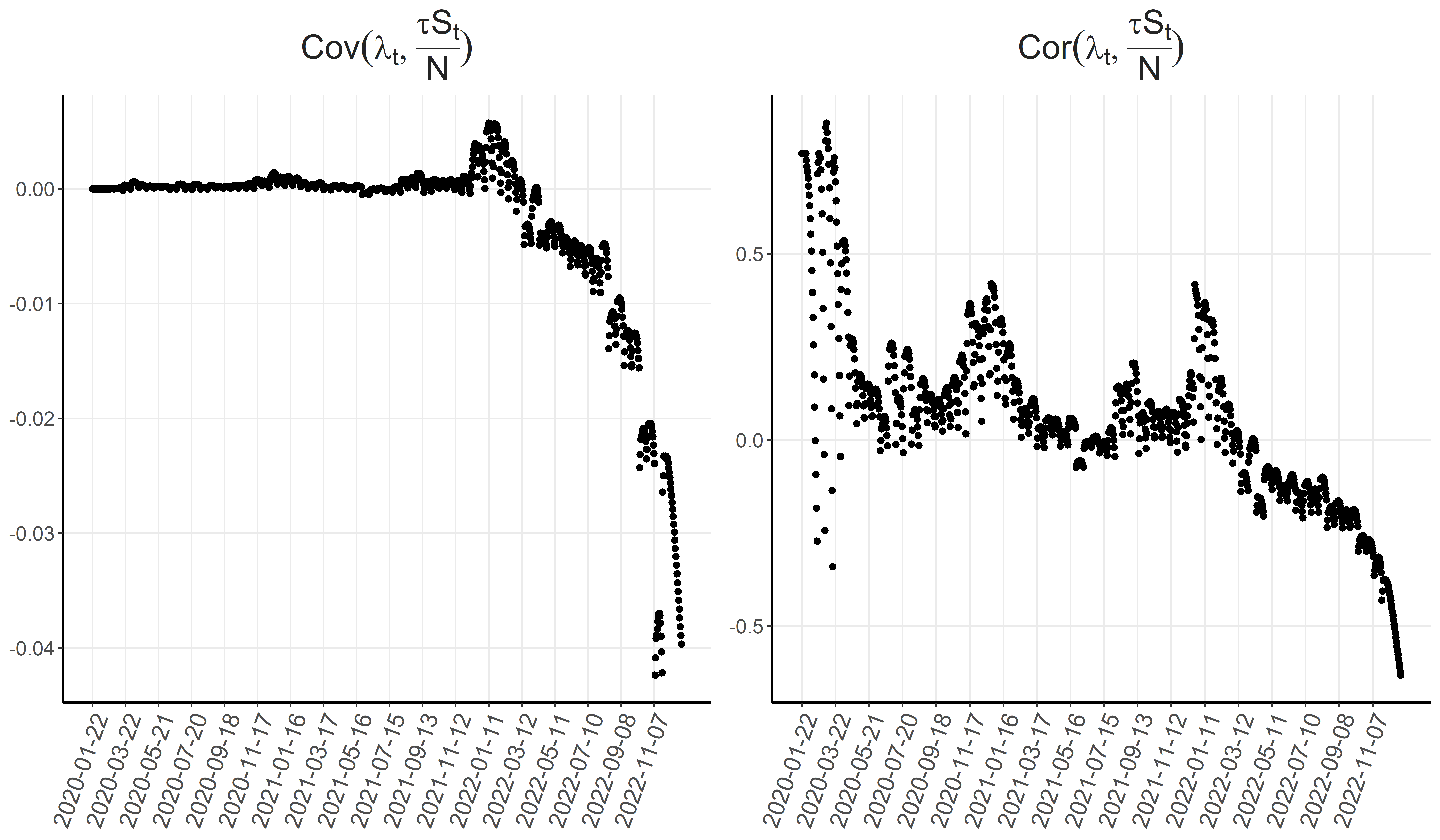}
\end{center}
\caption{Daily estimates of the covariance (left) and correlation (right) between $\lambda_t$ and $\tau S_t/N$ for USA.}
\label{fig:covsrs_us}
\end{figure}

Defining the end of the acute phase of the epidemic is a non-standard problem. The definition of the reproduction number $R_t=\lambda_t \cdot \tau \cdot S_t/N$ implies that at the early stage of the epidemic  $R_t$ is proportional to the infection rate. The covariance of the infection rate and the scaling factor $\tau S_t/N$ offers an insight to when those two diverge (see Figure \ref{fig:covsrs_us}) whence the covariance becomes negative. An additional version of the divergence between $\lambda_t$ and $R_t$ may be found in Supplementary Material H.

\section{Phase Plane Analysis}
\label{s:phasean}

In this Section we examine the model behaviour through the lens of dynamical non-linear systems. In particular, the non-pharmaceutical interventions are expressed in a novel way and a qualitative analysis of the epidemic is offered through the visual inspection and interpretation of certain plots. We introduce the main concepts in continuous time, before adapting to the discrete time used in this paper.

\subsection{Deterministic Epidemics on the Continuous Time Domain}
The system of coupled equations of $S$ and $I$ may be inspected focusing on the bivariate space $S\times I$ and the dynamic behaviour on this $SI$-plane, denoted as the set $V \subseteq \mathbb{R}_+^2$. Then, the standard SIR system can be examined as a vector field $\mathbf{F}$ on $V$. We define the \textit{natural epidemic flow} to be a curve $\bsigma(t)=\big(S(t),I(t)\big)$ that satisfies
\begin{equation*}
\mathbf{F\big(\bsigma(t)\big)}=\frac{dS}{dt} \cdot \ihat + \frac{dI}{dt} \cdot \jhat
\end{equation*}
Given an initial condition $\bsigma(0)$ the epidemic flow defines a trajectory on $V$, which we call the \textit{natural course of the epidemic}. The image of the realized transformation of time $t$ to the $\mathbb{R}_+^2$ plane $\bsigma:\mathbb{R}_+ \rightarrow \mathbb{R}_+^2$ (that is the realized trajectory) is a useful plane curve since it defines a path whose $x$ and $y$ coordinates represent the number of susceptible and infected individuals respectively at each time $t$. Thus, monitoring the path a particle follows in the epidemic flow, starting from the initial position $(S(0),I(0))$ can be an informative representation of the epidemic. 

The actual behaviour of the epidemic model of the present paper departs from the expected dynamics since the parameters change over time and the intervention measures interrupt the theoretical dynamics. The definition of the \textit{actual course of the epidemic} relates to a curve $\bgamma(t)$ on the $SI$-plane whose tangent vector at some points is different from the vector field at those points (in contrast to the natural course) and the departure is depicted in the ($\bgamma$, $\bsigma$) pair. Two measures of \textit{intervention effectiveness} can be constructed based upon the difference of the two courses. The first is based on the area between the positions under the actual and the natural course during some time interval $(a,b)$. Thus, we define
\begin{equation} \label{eq:vecdifC}
L_{a,b}=\int\limits_a^b \frac{||\bsigma(t)-\bgamma(t)||}{||\bsigma(t)||} dt
\end{equation}
where $\bsigma$ is the natural and $\bgamma$ the actual course. Although the natural course is unknown in practice, setting it to the current course and letting $\bgamma$ be the course under a hypothetical scenario where an intervention measure is taken leads to a measure of effectiveness quantification of this intervention. Large values of $L_{a,b}$ suggest large deviation from the initial course. The second measure is obtained by comparing the \textit{epidemic work} between the natural and actual course during the time interval $(a,b)$ with the epidemic work being defined as 
\begin{equation}
W_{a,b} = \int_a^b\left(\frac{dS}{dt}\right)^2+\left(\frac{dI}{dt}\right)^2dt \label{eq:SIwor} 
\end{equation}
i.e. the integrated squared speed
\begin{equation}
v(t) = \sqrt{\left(\frac{dS}{dt}\right)^2+\left(\frac{dI}{dt}\right)^2} \label{eq:SIspe}
\end{equation}
Lower speed is indicative of a better course, thus the larger the difference between the actual course and the one suggested by the intervention, the better. To this end, we define
\begin{equation} \label{eq:wordifC}
M_{a,b}=\frac{\displaystyle\int_\bsigma \mathbf{F} ds - \int_\bgamma \mathbf{F} dg}{\displaystyle\int_\bsigma \mathbf{F} ds}
\end{equation}
as the second intervention effectiveness measure with $M_{a,b}^{(A)}>M_{a,b}^{(B)}$ meaning that intervention A is to be preferred over intervention B.

Note that the simple SIR system can lead to a conserved quantity $Q(t)=S(t) + I(t) - \displaystyle\frac{1}{\lambda\tau} \log\big(S(t)\big)$, which characterizes this particular system. Inspecting $Q(t)$ at the start of the epidemic gives a measure of the deviation of the realised epidemic from the SIR formulation. Under SIR, $Q_0(t)$ is constant for initial values $(S_0,I_0,R_0)$, so $(Q_0^*(t)-Q_0(t))^2$ (where $Q_0^*(t)$ is obtained by a different model) measures the deviation from the SIR for the specific set of initial values $(S_0,I_0,R_0)$ for a specific time $t$. The absolute difference may also be used in this measure.

\subsection{Estimation on the Discrete Time Domain}

\begin{figure}
\begin{center}
\includegraphics[width=0.85\textwidth]{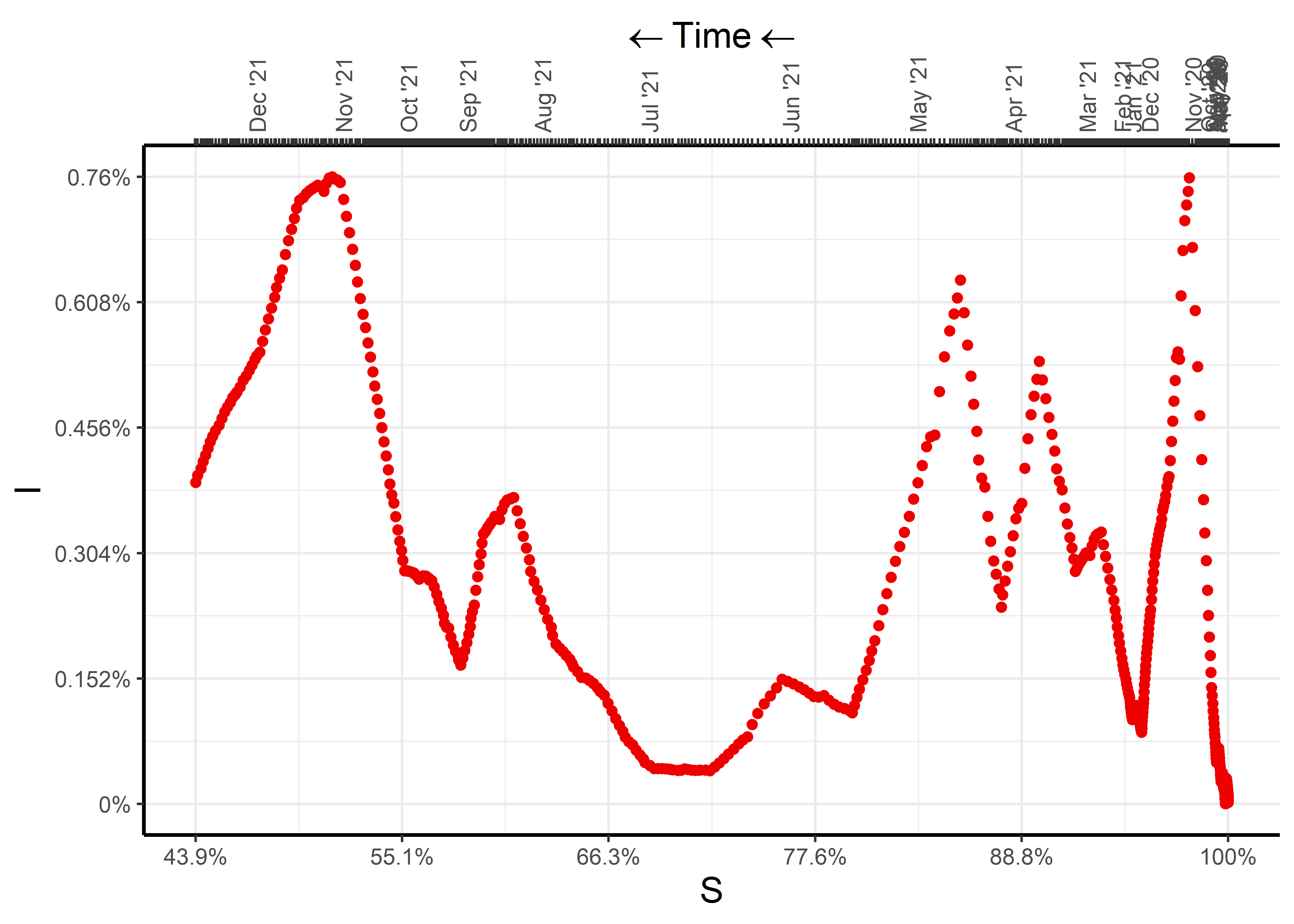}
\end{center}
\caption{The SEIR with vaccination and demography model estimates on the $SI$-plane.}
\label{fig:psseirvd}
\end{figure}

Since our models are defined in discrete time we develop here the discrete analogues and illustrate these concepts on the daily Covid-19 data from Greece. The SEIR model with vaccination and demography as proposed in this article provides estimates for the posterior medians of the susceptible and infectious individuals, which we plot in Figure \ref{fig:psseirvd} on the $SI$-plane as proportions of the total population. The red dots plot the estimated epidemic course $\hat{\bgamma}_t=(\hat{S}_t, \hat{I}_t)$, which is initiated at the bottom right corner $\hat{\bgamma}_0=(N_{GR}-\hat{C}_1,\hat{C}_1)$. We include a top axis whose ticks correspond to each $(\hat{S}_t,\hat{I}_t)$ point facilitating reading of the points of high and low speed (with respect to changes in $S$) given in (\ref{eq:SIspe}), which now reads
\begin{equation} \label{eq:SIspeD}
v_t=\sqrt{(S_{t+1}-S_t)^2+(I_{t+1}-I_t)^2}
\end{equation}
while the epidemic work given in (\ref{eq:SIwor}) can be obtained by
\begin{equation} \label{eq:SIvelD}
W_{a,b} = \sum\limits_{t=a}^{b-1}(S_{t+1}-S_t)^2 + (I_{t+1}-I_t)^2
\end{equation}
It is apparent that the epidemic at the end of 2021 was at the point $(S,I)=(0.4388,0.0039)$, so approximately 4746069 of the Greek population escaped infection, while 421234 was the active set of those infected at that time.

The intervention effectiveness for the time interval $(a,b)$ may be measured using
\begin{equation} \label{eq:vecdifD}
L_{a,b} = \sum\limits_{t=a}^b \sqrt{\frac{(S_t^{(n)}-S_t^{(a)})^2 + (I_t^{(n)}-I_t^{(a)})^2}{S_t^{(n)2}+I_t^{(n)2}}}
\end{equation}
and
\begin{equation} \label{eq:wordifD}
M_{a,b} = \frac{W_{a,b}^{(n)}-W_{a,b}^{(a)}}{W_{a,b}^{(n)}}
\end{equation}
where the superscripts $(n)$ and $(a)$ indicate natural and actual course respectively (or courses under the current and a new model).

\begin{figure}
\begin{center}
\includegraphics[width=\textwidth]{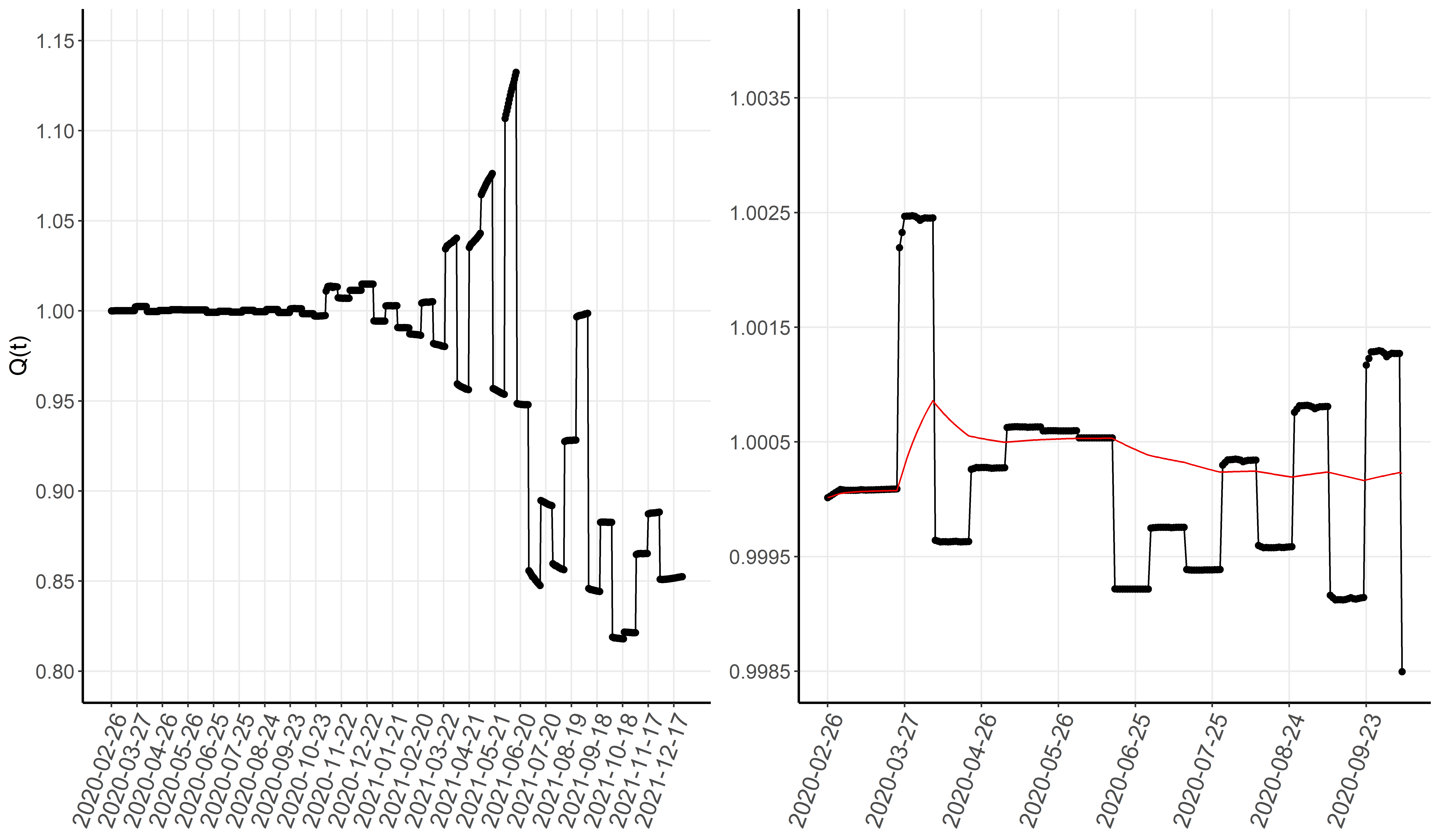}
\end{center}
\caption{Left: $Q_t$ for the $(S,I)$ values of Greece. There is a downward pattern after approximately 224 days. Right: Zoomed in picture of the first 225 days for better visualization of the random fluctuations of $Q_t$, albeit approximately constant in that interval. Red line is used for the ergodic mean of the points.}
\label{fig:Qseirvd12}
\end{figure}

We use the estimated $(S,I)$ quantities for Greece to examine the SIR assumption by checking the quantity $Q_t=S_t + I_t - \log S_t/(\lambda_t\tau)$. If the SIR assumption holds then $Q_t$ has to be roughly constant. In the left plot of Figure \ref{fig:Qseirvd12} a downward trend is apparent after the initial period suggesting a departure from the SIR assumption after about 224 days. A zoomed version for this time interval is depicted in the right plot of the same Figure. Due to the random fluctuations in $S$ and $I$ we also plot the ergodic mean (in red). The $Q_t$ values are around 1.000349 and this approach may serve as a goodness of fit test for the SIR formulation.

\section{Discussion}
\label{s:discuss}

This paper suggests that the first two years of the Covid-19 epidemic may reasonably be described by SEIR-type models which include information on demographic births and deaths as well as vaccination. Our inference is based upon the estimated total cases and the external validity of the results is inspected through the comparison with an independent dataset from the REACT-2 study in UK. The full SEIR model with vaccination and demography is tested removing some of its structural components in turn and comparisons are made using information criteria, the marginal likelihood and Bayes factors.

The proposed framework is developed based on publicly available data which are central to this work. Estimation is based on the NUTS variant of HMC which appeared to be the most reliable compared to variational Bayes and maximization of the un-normalized posterior via Simulated Annealing. The predictive ability of the mobility data is examined with or without dimension reduction and is shown to be relatively limited for accurate predictions and therefore for informing public health decisions (see Supplementary Material A, B, C, G).

A potential weakness of the proposed model is the reliance on the IFR. This is due to our focus upon publicly available data but our framework can accommodate additional evidence, like seroprevalence surveys or ICU data (e.g. \citealp{bergstrom2022bayesian}) in a straightforward manner. Prevalence data would give direct evidence on the IFR but they are not typically available for many countries and therefore are beyond the scope of this work. We found that informative priors on the IFR are preferable to fixing it to a point mass. The computational burden for training our models poses limits to certain short-term prediction exercises. However, one-week-ahead predictions are certainly feasible and often the most realistic if one wishes to avoid strong assumptions on the population behaviour.

The dynamics of the proposed epidemic models are examined on the phase plane of the susceptible and infectious individuals, offering estimates of effectiveness for control measures such as the non-pharmaceutical interventions adopted in the early phase of the pandemic. A simple goodness-of-fit type of assessment for the SIR assumption is proposed based on the time-invariance of the $Q_t$ quantity. Further work based on variants of $Q_t$ is required and this is the subject of future research.

\bibliographystyle{Chicago}
\bibliography{bibtex}

\end{document}